\documentclass[prl,twocolumn,showpacs,superscriptaddress,preprintnumbers]{revtex4-1}
\usepackage{amsfonts, amsmath, amssymb, bm, bbm, graphicx, epsfig, epstopdf}
\usepackage{dcolumn}
\usepackage{textcomp}
\usepackage{hyperref}

\begin{document}

\title{Entanglement over global distances via quantum repeaters with satellite links}

\author{K. Boone}
\affiliation{Institute for Quantum Science and Technology and
Department of Physics and Astronomy, University of Calgary,
Calgary T2N 1N4, Alberta, Canada}

\author{J.-P. Bourgoin}
\affiliation{Institute for Quantum Computing, University of Waterloo, Waterloo, ON N2L 3G1, Canada}

\author{E. Meyer-Scott}
\affiliation{Institute for Quantum Computing, University of Waterloo, Waterloo, ON N2L 3G1, Canada}

\author{K. Heshami}
\email{khabat.heshami@nrc.gc.ca}
\affiliation{Institute for Quantum Science and Technology and
Department of Physics and Astronomy, University of Calgary,
Calgary T2N 1N4, Alberta, Canada}
\affiliation{National Research Council of Canada, 100 Sussex Drive, Ottawa, Ontario, K1A 0R6, Canada}

\author{T. Jennewein}
\email{thomas.jennewein@uwaterloo.ca}
\affiliation{Institute for Quantum Computing, University of Waterloo, Waterloo, ON N2L 3G1, Canada}
\affiliation{Quantum Information Science Program, Canadian Institute for Advanced Research, Toronto, ON, Canada}

\author{C. Simon}
\email{csimo@ucalgary.ca}
\affiliation{Institute for Quantum Science and Technology and
Department of Physics and Astronomy, University of Calgary,
Calgary T2N 1N4, Alberta, Canada}

\begin{abstract}
We study entanglement creation over global distances based on a quantum repeater architecture that uses low-earth orbit satellites equipped with entangled photon sources, as well as ground stations equipped with quantum non-demolition detectors and quantum memories. We show that this approach allows entanglement creation at viable rates over distances that are inaccessible via direct transmission through optical fibers or even from very distant satellites.
\end{abstract}
\maketitle

Over the last few decades the distribution of quantum entanglement has progressed from tabletop experiments to distances of over one hundred kilometers \cite{entdist}. Will it be possible to create entanglement over global distances? This is interesting from a fundamental point of view, but also from the perspective of trying to create a global ``quantum internet'' \cite{kimble}. In the context of quantum cryptography, it would enable secure global communication without having to rely on any trusted nodes \cite{jennewein}, as entanglement is the foundation for device-independent quantum key distribution \cite{diqkd}. It would also be useful for global clock networks \cite{lukin-clocks} and for very long baseline telescopes \cite{gottesman}.

Modern classical telecommunication relies on optical fibers. Unfortunately the direct transmission of photons through fibers is not practical for quantum communication over global distances because losses are too high. The best available fibers have a loss of 0.15 dB/km at the optimal wavelength. This means, for example, that the time to distribute one entangled photon pair over 2000 km with a 1 GHz source exceeds the age of the universe.

Two alternative approaches to try to overcome this problem are currently being pursued in parallel, namely fiber-based quantum repeaters and direct satellite links. Conventional quantum repeaters rely on first creating and storing entanglement for elementary links, then extending the distance of entanglement by entanglement swapping \cite{repeaters,repeaters-RMP}. Based on the experimental and theoretical progress in this area over the last few years, it is plausible that this approach will make it possible to extend the distance of entanglement distribution significantly beyond what is possible with direct transmission through optical fibers \cite{repeaters-RMP,repeaters-recent,krovi}. However, truly global distances are still very difficult to envision for repeaters based on fiber links. This is true also for related approaches based on quantum error correction \cite{errorcorr}, which tend to require repeater stations that are only a few kilometers apart.

\begin{figure}
\scalebox{0.46}{\includegraphics*[viewport=40 395 580 600]{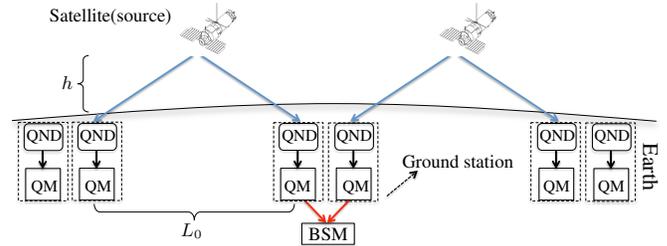}}
\caption{(Color online) Proposed quantum repeater architecture with satellite links. Each elementary link (of length $L_0$) consists of an entangled photon pair source on a low-earth orbit satellite (at height $h$), and two ground stations consisting of quantum non-demolition (QND) measurement devices and quantum memories (QM). The successful transmission of entangled photons to each ground station is heralded by the QND devices, which detect the presence of a photon non-destructively and without revealing its quantum state. The entanglement is then stored in the memories until information about successful entanglement creation in two neighboring links is received. Then the entanglement can be extended by entanglement swapping based on a Bell state measurement (BSM). Figure 2 shows that four to eight such links are sufficient for spanning global distances. }\label{architecture}
\end{figure}

The use of satellite links for quantum communication is also being pursued very actively. There has been a lot of progress in terms of feasibility studies \cite{Buttler,Kurtsiefer,Rarity,Bonato,PanSatExp,Bourgoin,Ling,SatExp}. The launch of the first satellite carrying an entangled pair source has been announced for 2016 \cite{xin}. The advantage of quantum communication via satellites is that transmission loss is dominated by diffraction rather than absorption and thus scales much more favorably with distance. For example, consider a pair source on a satellite at a height of 1000 km. For realistic assumptions (such as telescope size, see below), the combined transmission loss for the photon pair for a 2,000 km ground station distance is only of order 40~dB. This should be contrasted with 300 dB for a fiber link of the same length. However, global distances are still challenging even for satellite links. Direct transmission from low-earth orbit (LEO) satellites, i.e. those below the Van Allen radiation belt, or up to about 2000 km in height, no longer works. Even before the Earth gets in the way, the loss becomes forbidding for very grazing incidence due to long propagation distance in air. One possible solution is to use satellites that are much further away, but this comes at significant cost, as satellites have to be much more robust to shield them from radiation. Moreover the greatest ground distances, approaching 20,000~km (i.e. half the Earth's circumference), are out of range even for very distant satellites.

Here we propose to combine the two approaches discussed above. We study quantum repeaters based on LEO satellite links, as illustrated in Figure 1. The satellites just need to be equipped with entangled pair sources, while the more complex components, such as quantum memories and quantum non-demolition (QND) detectors, are on the ground and can be further developed even after the satellites are launched. An important difference between satellite and fiber-based links is that the satellite-based links are active only during each time period when the satellite is visible from both ground stations (the ``flyby time'' $T_{FB}$). For currently realistic quantum memory lifetimes all satellite links in Figure 1 have to be active simultaneously, which implies that our architecture requires a number of satellites equal to the number of links. However, our results show that four to eight links are sufficient to span global distances.

Figure 2 compares the expected entanglement distribution rates per day for repeater architectures with LEO satellites to those achievable by direct transmission from more distant satellites. It is important to make the comparison on a per day basis since the flyby times and periods are different for satellites at different heights. Our results suggest that approach based on repeaters with LEO satellite links is attractive for all but the shortest distances and is the only way to create entanglement for the longest distances. We now describe the assumptions and requirements underlying these results in some detail.

\begin{figure}
\scalebox{0.48}{\includegraphics*[viewport=30 183 530 590]{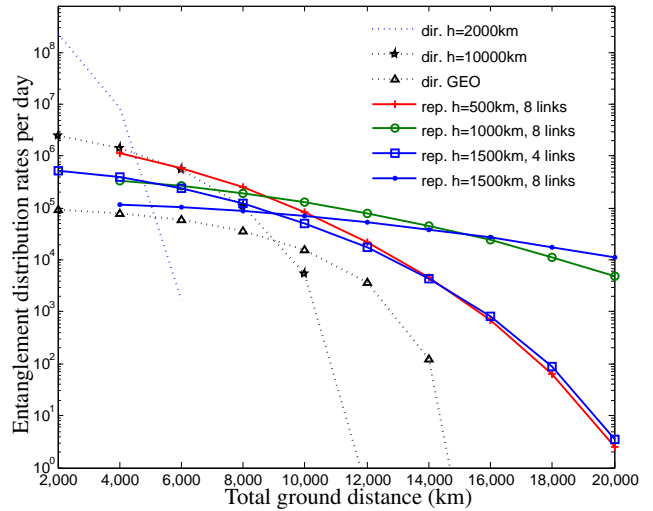}}\caption{(Color online) Entanglement distribution rates per day as a function of ground distance for quantum repeaters with LEO satellites (solid lines) at heights $h$=500, 1,000, 1,500~km, compared to direct transmission (dotted lines) from satellites at heights $h$=2,000 and 10,000~km and from a geostationary satellite.}
\label{rates}
\end{figure}

One key ingredient for our analysis is the calculation of the probability for a pair of photons that are emitted from a satellite at height $h$ to be successfully transmitted to the ground. Our approach, which is based on Ref. \cite{Bourgoin}, takes into account diffraction, pointing error and atmospheric transmittance. In Figure 2 we assume a satellite transmitter size of 50~cm and ground telescope size of 1~m. For the quantum repeater scenarios we assumed a pair source that emits photons at 580~nm, which is motivated by our choice of quantum memory material (Eu-doped yttrium orthosilicate, see below). For direct transmission we assumed a wavelength of 670~nm (470~nm) for $h=2000$ km ($h=10,000$ km and geostationary), which results in optimal transmittance \cite{Bourgoin}. We include a satellite pointing error of 0.5~$\mu$rad and assume ground stations at rural atmosphere at sea level; see the supplementary information \cite{supplementary} for more details. We assume that frequency shifts due to relativistic and gravitational effects \cite{freqshifts} are compensated (e.g. by acousto-optic modulators on the ground). Timing jitter due to turbulence in the atmosphere is negligible for the relatively long pulses that we are considering \cite{kral}.

For the repeater scenarios, we have assumed a pair source with a repetition rate of 10~MHz. This value is motivated primarily by the expected memory bandwidth for our choice of material, see below. In contrast, we assume a 1~GHz repetition rate for direct transmission.
In each case the source could e.g. be a deterministic pair source based on a quantum dot in microcavity \cite{qdots}. However, simpler implementations are possible based on parametric down-conversion sources with a small pair creation probability per pulse (below 0.01) \cite{PDC}, in order to avoid errors due to multi-pair emissions. If one aims to achieve the same effective rate in this way, the underlying repetition rate (and hence memory bandwidth, in the repeater scenario) has to be increased correspondingly. Memory bandwidths up to 1 GHz have already been achieved in rare-earth doped materials (e.g. in Tm:LiNbO \cite{Saglamyurek}), but not yet in combination with long storage times. We have not assumed any frequency multiplexing, neither for repeaters nor for direct transmission. This could be used to boost rates in both scenarios, at the expense of more complex sources on the satellites.

The rates for the repeaters are calculated as in Ref. \cite{repeaters-RMP}, assuming a ``nested'' approach. That is, entanglement is first created and stored at the level of the elementary links. Then links are connected in a hierarchical fashion, forming links of length two, four etc. For convenience let us define the average
probability of a pair reaching the ground stations during one flyby of the satellite as $P_0^{\text{avg}}=\frac{\int\eta^{(2)}_{\text{tr}}(t)dt}{T_{\text{FB}}}$, where $\eta^{(2)}_{\text{tr}}(t)$ is the time-dependent combined two-photon transmission, and $T_{\text{FB}}$ is the flyby time of the satellite \cite{supplementary}. The probability of successfully creating, transmitting  and storing an entangled pair over one elementary link is $P_{\text{EG}}=\eta_{\text{s}} P_0^{\text{avg}} \eta_{\text{q}}^2 \eta_{\text{w}}^2$, where $\eta_{\text{s}}$, $\eta_{\text{q}}$ and $\eta_{\text{w}}$ are source, QND detector and memory write efficiencies. Entanglement swapping relies on Bell-state measurements (BSM). In our scheme, a successful BSM requires successful readout of two photons from neighboring quantum memories with the efficiency of $\eta_{\text{r}}^2$ and two single-photon detections with $\eta_{\text{d}}^2$ efficiency. Here $\eta_{\text{r}}$ and $\eta_{\text{d}}$ are memory readout and detector efficiencies. This gives the entanglement swapping efficiency of $P_{\text{ES}}=\frac{\eta_{\text{r}}^2\eta_{\text{d}}^2}{2}$, where the factor of 1/2 is due to limited success probability of the BSM using linear optics with ancillary vacuum modes \cite{LinearBSM}. Higher success probabilities are possible in principle using ancillary photons \cite{Grice, vanLoock14}. For a repeater composed of $2^n$ links, the number of entangled pairs created during one flyby is given by $R_{\text{s}} T_{\text{FB}}P_{\text{EG}} (\frac{2}{3}P_{\text{ES}})^n$, where $R_{\text{s}}$ is the source rate. The factors of 2/3 take into account the fact that entanglement has to be created in two neighboring links before entanglement swapping can proceed \cite{repeaters-RMP}. In Fig.~\ref{rates}, we assumed that the different elements including source, QND detector, quantum memory (write and read) and detectors have efficiencies of 0.9. The effect of inefficiencies in these elements on the total entanglement distribution rate is shown in Figure 3 and in the supplementary information \cite{supplementary}. Significantly higher efficiencies have already been achieved for single-photon detectors \cite{nam}. We now discuss the other elements in more detail.

\begin{figure}
\scalebox{0.58}{\includegraphics*[viewport=65 200 460 775]{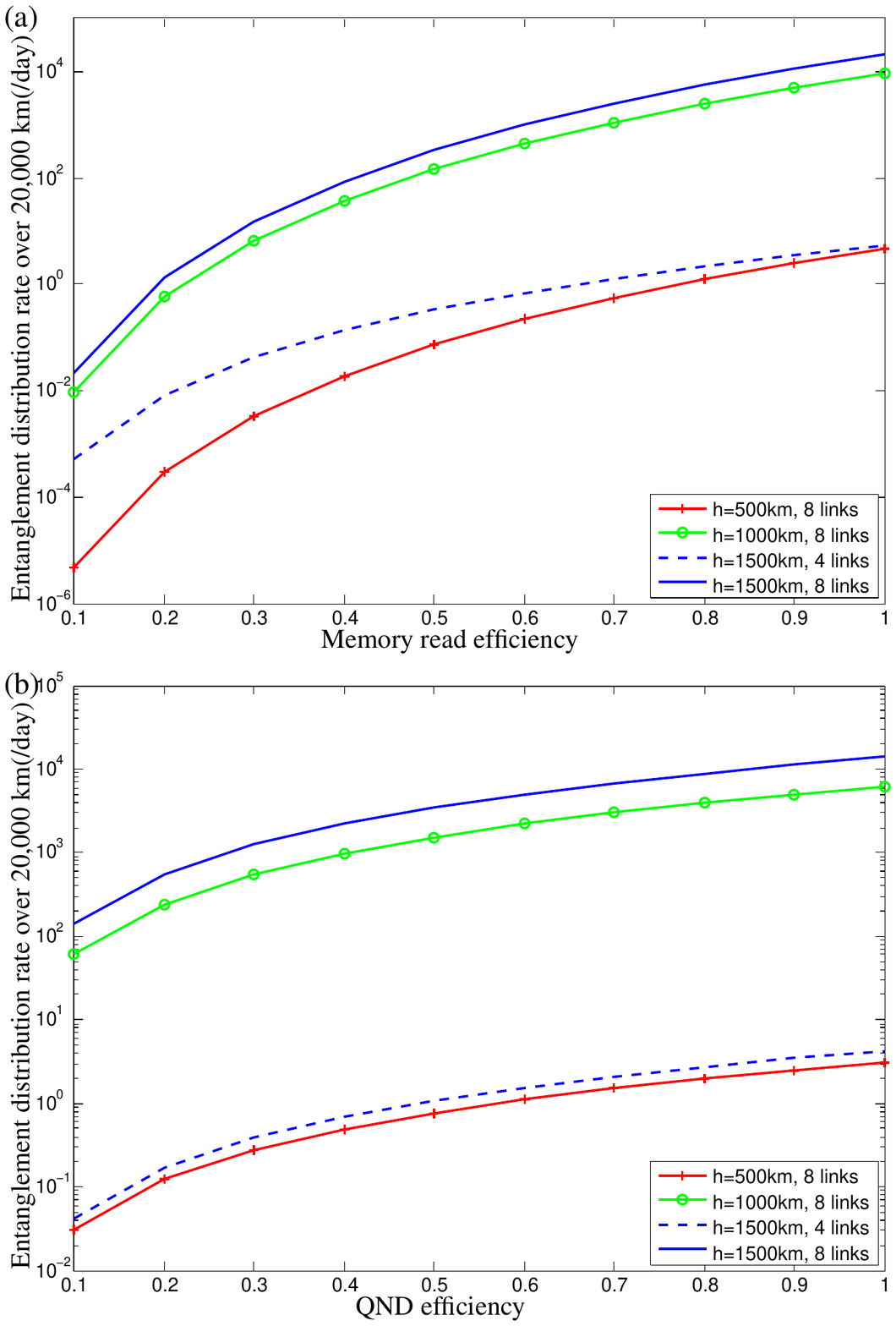}}
\caption{(Color online) Impact of inefficiencies in various elements on the entanglement distribution rate over 20,000 km for the repeater protocols shown in Figure 2. (a) Effect of the memory read efficiency. The detector efficiency has the same effect. (b) Effect of QND detector efficiency. Memory write efficiency and source efficiency have similar effects, see also the supplementary information \cite{supplementary}. The repeater protocol is more sensitive to memory read and detector efficiency than to that of the other components, because the former efficiencies intervene in each entanglement swapping step, whereas the latter only intervene in the entanglement creation in the elementary links.}
\label{inefficiency}
\end{figure}

Quantum memories for photons have been implemented in a range of physical systems \cite{memories}. Memories based on rare-earth ion doped crystals \cite{RE-memories} are particularly attractive for our purpose because of their potential for highly multi-mode storage, e.g. using the atomic frequency comb (AFC) protocol \cite{AFC}. This is important because the quantum memories in each ground station will be exposed to a large number of photons, $N_{\text{mod}}=R_{\text{s}}\eta_{\text{s}} \eta^{(1)}_{\text{tr},\text{max}}\frac{L_0}{c}$, before receiving the classical signals from the other end of each link that make it possible to decide which photons are part of an entangled pair and should thus be kept for entanglement swapping. Here $\eta^{(1)}_{\text{tr},\text{max}}$ denotes the maximum value of the single-photon transmission during one flyby. For the quantum repeater scenarios in Fig. \ref{rates}, multimode storage of $10^3-10^4$ photons is required according to the above formula (depending on satellite height and number of links). A single AFC type quantum memory based on Eu-doped yttrium ortho silicate (YSO) should be able to store $10^2-10^3$ photons in distinct temporal modes \cite{AFC}. Having several waveguides or using multiple locations on the same crystal makes the storage of a total of $10^4$ photons in distinct modes in a single crystal plausible. Our protocol also requires storage times of the order of the total communication time $L/c$, where $L$ is the total distance, which corresponds to 67 ms for 20,000 km. Such long storage times should be achievable by transferring the optical memory excitations to ground spin states \cite{Afzelius10,EuQM}. Note that storage times up to one minute have already been achieved in rare-earth doped crystals in the classical regime \cite{Halfmann}. The requirement of transferring the excitation to the ground state limits the repetition rate of the photon source as the bandwidth of the photons must be smaller than energy spacing between the ground spin states. The 10 MHz bandwidth assumed in Figure 2 is compatible with the ground level separations of Eu:YSO, which are of order 100 MHz \cite{EuQM}. High memory efficiencies, which are also important for our purpose (see Figure 3), can be achieved in rare-earth doped crystals with the help of optical cavities \cite{AFC-cavity}.

Our scheme also requires QND detection of the photonic qubits. QND measurement of photons has recently been demonstrated using a single atom in a cavity \cite{RempeQND}. The cross-Kerr effect induced by the AC-Stark shift in atomic ensembles also provides the possibility to realize QND measurement of photons. In Ref. \cite{Gaeta}, 0.5~mrad cross-phase shift per photon has been shown using a hot atomic vapor inside a hollow-core photonic crystal fiber, which should already allow a QND measurement of the photon number \cite{haus}. Here we also require the QND measurement to be insensitive to the photonic qubit state. For example, if photon pairs with polarization entanglement are to be detected, the probe field must interact with both polarization modes. A simpler implementation of the QND detection of photonic qubits is possible for time-bin qubits based on the AC-Stark shift in combination with quantum storage because the phase shift imparted to the stored probe beam is not sensitive to the precise timing of the signal photon propagating through the ensemble \cite{Heshami14}. This approach should also make it possible to integrate the QND detector with the quantum memory, e.g. a rare-earth doped waveguide \cite{Saglamyurek}. Another possibility is to use a heralded qubit amplifier based on linear optics and a deterministic pair source \cite{Meyer-Scott}. This achieves a QND detection efficiency of up to 0.5. Compared to the assumed QND detection efficiency of 0.9 in Fig. \ref{rates}, this would reduce the entanglement distribution rate by approximately a factor of 5 (see Fig. \ref{inefficiency}).

We discuss the effect of background photons in the supplementary information \cite{supplementary}. Due to the narrow bandwidth of the photons used, which allows the use of correspondingly narrow filters, the background noise is completely negligible at nighttime. For the bright daytime sky, the related errors are below the 1\% level for the repeater scenarios in Figure 2. In contrast, for direct transmission from a GEO satellite the background noise becomes dominant for distances beyond 12,000 km.

Due to turbulence, which disturbs the wavefront of the photons, the use of adaptive optics is likely to be required in order to couple the photons into single-mode waveguides for QND detection and quantum storage. This also makes it possible to consider alternative repeater scenarios based on uplinks \cite{uplink-repeaters}, which are significantly more lossy than downlinks without adaptive optics. See the supplementary information \cite{supplementary} for more discussion of this possibility.

We only performed a simple rate calculation for the proposed repeater architecture. A more sophisticated analysis would characterize the fidelity of the distributed quantum state and extract a key rate for quantum key distribution applications \cite{krovi}. However, assuming low noise levels in all components, and given the fact that we only consider small numbers of repeater links, the present estimates should give a reasonably accurate picture of achievable key rates.

We have argued that quantum repeaters based on LEO satellite links are a viable approach to global quantum communication. Our proposed scheme relies on realistic advances in quantum memories and quantum non-demolition measurements and only requires a moderate number of satellites equipped with entangled photon pair sources. Ultimately global quantum repeater networks will likely combine satellite links for very long distances with fiber links for short and intermediate distances.

{\it Acknowledgments.} The authors would like to thank W. Tittel for valuable discussions. This work was supported by NSERC, AITF, DARPA,
Ontario Ministry of Research and Innovation, ERA, Industry Canada, CIFAR, CFI, and FEDDEV Ontario.

\newpage

\widetext
\begin{center}
\textbf{\large Supplemental Material for ``Global quantum communication with satellites and quantum repeaters"}
\end{center}

\setcounter{table}{0}
\setcounter{equation}{0}
\setcounter{figure}{0}
\makeatletter
\renewcommand{\bibnumfmt}[1]{[SI#1]}
\renewcommand{\citenumfont}[1]{SI#1}
\renewcommand{\figurename}[1]{FIG. SI#1}

\section{Transmission Calculation}
Our transmission calculations for the satellite links take into account diffraction, pointing error and atmospheric transmittance~\cite{JP-NJP13}. Diffraction is modeled using the Rayleigh-Sommerfeld formula~\cite{G96}. Pointing error, which arises due to jitter in the telescope and imprecision in the tracking system, will cause the beam to wander with a Gaussian distribution. We model the average combined effect of diffraction and pointing error by taking a two-dimensional convolution~\cite{B09} of the diffraction profile with the Gaussian distribution of the pointing error. Finally the atmospheric transmittance, characterizing the probability of light to be scattered or absorbed by molecules in the atmosphere, is modeled using the MODTRAN 5 software package. The transmission calculation is performed for every point (at 10~s intervals) where the satellite is in view of the ground station.

\begin{figure}[ht]
\scalebox{0.9}{\includegraphics*[viewport=15 440 590 650]{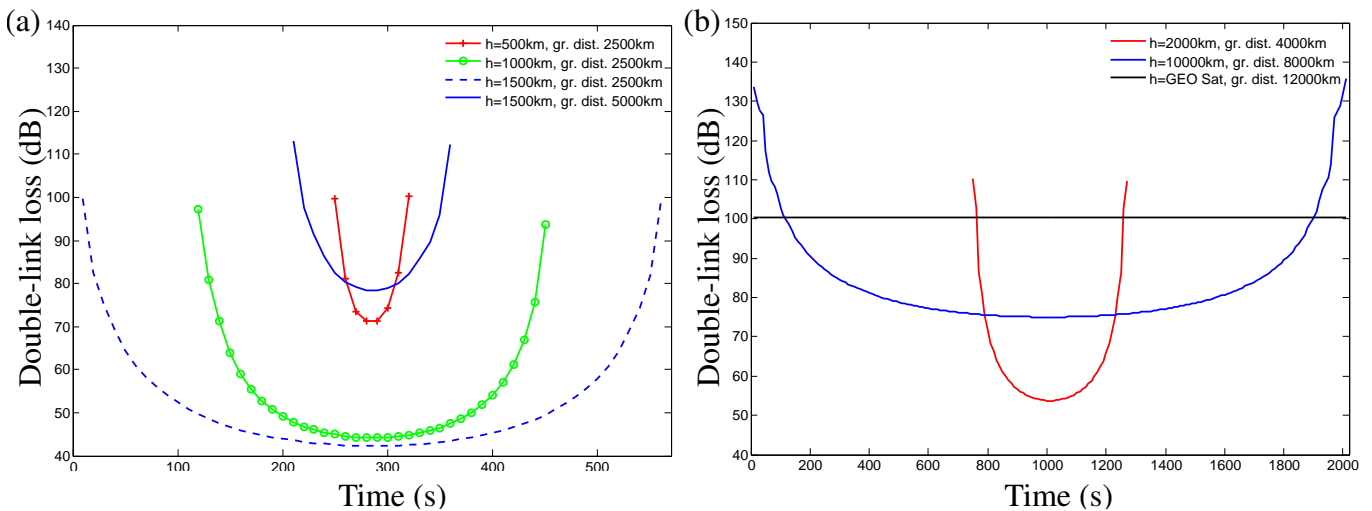}}
\caption{(Color online) (a) Double-link loss for the LEO satellites used in our proposed quantum repeater architecture for representative ground distances. (b) Double-link loss for the high altitude satellites that are considered in the direct transmission scenarios.}\label{DLLRep}
\end{figure}

In Fig. SI 1 we show results for the double-link loss for the LEO and higher altitude satellites used in Fig. 2 in the main text. This allows us to find the average transmission and the flyby time for each configuration.

\section{Orbit analysis}
The period of the satellite is given by $T=2\pi/(\omega\pm2\pi/T_{\text{Earth}})$, where $\omega=\sqrt{GM/(R_{\text{e}}+h)^3}$ and $T_{\text{Earth}}$ is the rotation period of the Earth. $R_{\text{e}}$ and $M$ are the Earth's radius and mass, $h$ is the satellite's height and $G$ is the gravitational constant. Here we assumed a simple circular orbit on the equatorial plane with no inclination. For all orbits except GEO (which by definition must move in the same direction as the Earth's rotation), the overall performance is independent of the direction of the satellite. If the direction is the same as the Earth's rotation, the orbit will have longer but less frequent passes compared to an orbit with the opposite direction. In both cases the average loss and average contact time per day remains the same. For our analysis we chose the direction to be opposite of the rotation of the Earth, allowing more frequent key exchange at the cost of key length.

\section{Inefficiencies}
 We studied the dependence of the entanglement distribution rate over 20,000~km with respect to the efficiency of the different elements of our proposed architecture. As it can be seen in Fig. 3 in the main text and in Fig. SI 2, the entanglement distribution rate has a similar dependence on source, QND detection and memory write efficiency. This is because these elements contribute similarly to the entanglement generation rate. Detector and memory read efficiency have the same impact on the rate as they have the same contribution to the entanglement swapping success probability.
\begin{figure}[h]
\centering\scalebox{0.85}{\includegraphics*[viewport=0 150 600 320]{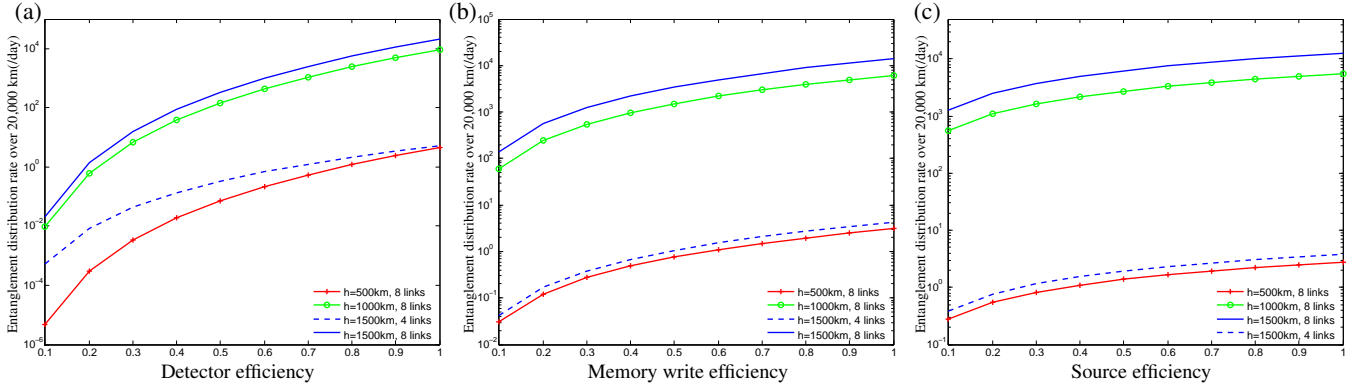}}\caption{(Color online) Entanglement distribution rate over 20,000~km vs. source, memory write and detector efficiency. All other efficiencies are set equal to 0.9.}
\label{Suppineffic}
\end{figure}

\section{Background Light}
A detailed nighttime background calculation \cite{JP-NJP13} showed a background on the order of 10$^2$/s with a 1~m receiver diameter, 1~nm filter and a 50~$\mu$rad field of view. For our quantum repeater scenario we have assumed a bandwidth of 10 MHz, allowing us to use a filter of that width (approximately 10$^{-5}$~nm). In addition the field of view of 50~$\mu$rad is a conservative estimate that can be reduced, reducing the background counts further by approximately 2 orders of magnitude. This leads to a nighttime background count on the order of 10$^{-4}$/s, which has negligible impact. The daylight background contribution has been studied in Ref.~\cite{LosAlamos1-supp}. They found the expected background counts to be on the order of 10$^{22}$/s/m$^2$/$\mu$m/str corresponding to around 100/s for a 1~m receiver diameter, 10~MHz filter and 10~$\mu$rad field of view.

The primary errors due to background light are caused by a noise photon arriving at one side in coincidence with a single photon from the source at the other side of an elementary link, which is wrongly interpreted as an entangled photon pair reaching the ground stations.
This coincidence probability is given by $P_{\text{noise}} P_{\text{single}}$, where $P_{\text{noise}}=R_{\text{noise}}T$ and $P_{\text{single}}=R_{\text{s}}\eta_{\text{single}}^{\text{max}}T$. Here, $R_{\text{noise}}$($R_{\text{s}}$) is the noise (source) rate, $T$ is the coincidence time window (which can be set to the inverse of the source rate) and $\eta_{\text{single}}^{\text{max}}$ is the maximum single-link transmission from the satellite to the ground. Comparing this coincidence probability (for daytime noise) with the probability of receiving entangled photon pairs for different LEO satellites shows that only about 1\% error occurs for our quantum repeater scheme. However, for the direct transmission scenario based on using a  geostationary satellite the noise will become dominant for distances beyond 12,000~km.

\section{Up-link scenario}
Using adaptive optics one can in principle achieve up-link losses approaching those of down-links. Therefore, a more conventional quantum repeater scheme with up-links (ground to satellite links) will become feasible \cite{rideout}. In such a scheme, pair sources and quantum memories remain on the ground and beam splitters and detectors for Bell state measurement are placed on the satellite. In this case the scheme does not require QND detectors. As the pair source is adjacent to the quantum memory at each node, photons from the pair source can be stored into quantum memories on the ground with ideally no loss. However, this means that in order to operate the source at its maximum possible rate this scheme would require approximately two orders of magnitude larger memory mode capacity.

\end{document}